\documentclass[aps,prd,twocolumn,groupedaddress,amssymb,eqsecnum,showpacs,epsfig]{revtex4}
\usepackage{graphicx}

\usepackage{bm}
\usepackage{latexsym}
\usepackage{amssymb}
\usepackage{amsfonts}
\usepackage{amsmath}
\usepackage{url}
\PassOptionsToPackage{hyphens}{url}
\usepackage[hidelinks]{hyperref}
\usepackage{epsfig}

\def \lleq {\lower0.9ex\hbox{ $\buildrel < \over \sim$} ~}
\def \ggeq {\lower0.9ex\hbox{ $\buildrel > \over \sim$} ~}

\def \beq  {\begin{equation}}
\def \eeq  {\end{equation}}
\def \ber  {\begin{eqnarray}}
\def \eer  {\end{eqnarray}}

\newcommand{\newc}{\newcommand}

\newc{\diag}{\mathop{\mathrm{diag}}}
\newc{\be}{\begin{equation}}
\newc{\ee}{\end{equation}}
\newc{\ba}{\begin{eqnarray}}
\newc{\ea}{\end{eqnarray}}
\newc{\bea}{\begin{eqnarray*}}
\newc{\eea}{\end{eqnarray*}}
\newc{\D}{\partial}
\newc{\ie}{{\it i.e.} }
\newc{\eg}{{\it e.g.} }
\newc{\etc}{{\it etc.} }
\newc{\etal}{{\it et al.}}
\newc{\lcdm }{$\Lambda$CDM }
\newcommand{\nn}{\nonumber}
\newc{\ra}{\rightarrow}
\newc{\lra}{\leftrightarrow}
\newc{\lsim}{\buildrel{<}\over{\sim}}
\newc{\gsim}{\buildrel{>}\over{\sim}}

\begin{document}

\title{Scalar-Tensor Quintessence with a linear potential:\\Avoiding the Big Crunch cosmic doomsday}
\author{A. Lykkas and L.~Perivolaropoulos}
 \affiliation{Department of Physics, University of Ioannina, Greece}
\date{\today}

\begin{abstract}
All quintessence potentials that are either monotonic with negative interval or have a minimum at negative values of the potential, generically predict a future collapse of the scale factor to a ``doomsday'' singularity. We show that this doomsday is generically avoided in models with a proper non-minimal coupling of the quintessence scalar field to the curvature scalar $R$. For simplicity we consider linear quintessence potential $V=-s\phi$ and linear non-minimal coupling $F=1-\lambda \phi$. However our result is generic and is due to the fact that the non-minimal coupling modifies the effective potential that determines the dynamics of the scalar field. Thus for each positive value of the parameter $s$ we find a critical value $\lambda_{crit}(s)$ such that for $\lambda>\lambda_{crit}(s)$ the negative potential energy does not dominate the universe and the cosmic doomsday Big Crunch singularity is avoided because the scalar field eventually rolls up its potential. We find that $\lambda_{crit}(s)$ increases approximately linearly with $s$. For $\lambda>\lambda_{crit}(s)$ the potential energy of the scalar field becomes positive and it eventually dominates while the dark energy equation of state parameter tends to $w=-1$ leading to a deSitter Universe.
\end{abstract}
\pacs{98.80.Es,98.65.Dx,98.62.Sb}
\maketitle
\section{Introduction}
Quintessence models where the potential takes negative values for a range of scalar field values are generic in a variety of theoretical models including $N=2,4,8$ gauged supergravity \cite{Kallosh:2001gr,Linde:2001ae}, brane cosmology \cite{Felder:2001da} and cyclic universe models \cite{Steinhardt:2001st,Steinhardt:2001vw}. The cosmological evolution in models involving such potentials has been studied extensively in the context of General Relativity \cite{Felder:2002jk,dooms-Kallosh:2003bq,dooms-Garriga:2003nm,lin-neg-Perivolaropoulos:2004yr,dooms-Wang:2004nm}.

It is well known that quintessence models with scalar field potentials that get negative for a range of field values generically predict the collapse of the cosmic scale factor to a Big Crunch singularity \cite{Felder:2002jk,dooms-Kallosh:2003bq,dooms-Garriga:2003nm,lin-neg-Perivolaropoulos:2004yr,dooms-Wang:2004nm} at a future cosmological time (cosmic doomsday). This collapse is due to the eventual scalar field evolution towards negative values of its potential where the gravity of the field is attractive. Such a behavior is generic for all quintessence models where the scalar field potential is not strictly positive but it has been studied for monotonic linear potentials of the form\cite{Felder:2002jk,dooms-Kallosh:2003bq,dooms-Garriga:2003nm,lin-neg-Perivolaropoulos:2004yr,dooms-Wang:2004nm}
\be
V(\phi)=-s\phi, \label{linpot-def}
\ee
where $s$ is a parameter. This class of potentials have interesting properties including a possible solution of the cosmic coincidence problem\cite{coincid-lin-Avelino:2004vy}. This behavior has been shown to be generic even in models beyond General Relativity where a non-minimal coupling of the scalar field to  matter is present \cite{Giambo:2014jfa}. It is therefore interesting to investigate the generic nature of this singularity in other modified gravity models.

The consideration of scalar-tensor quintessence models generically changes the dynamical evolution of the scalar field. In this case the evolution is determined by both the scalar field potential $V(\phi)$ and by the non-minimal coupling to gravity $F(\phi)$\cite{scal-tens-cosmo-EspositoFarese:2000ij,scal-tens-quint-Capozziello:2007iu,scal-tens-cross-Perivolaropoulos:2005yv,
scal-tens-rec-Boisseau:2000pr,scal-tens-quint-Torres:2002pe,scal-tens-dyn-Carloni:2007eu}. Thus it is possible that the cosmic doomsday singularity may be avoided in this class of models for a proper choice of the non-minimal coupling. This issue is investigated in the present study. 

In the next section we review the cosmological dynamics in the context of scalar tensor theories focusing on a flat cosmological model with a non-minimal coupling that is linear in the scalar field. We derive the cosmological dynamical equations for the homogeneous scalar field and the scale factor and present them in a rescaled form appropriate for numerical solution. In section III we solve the cosmological equations numerically and demonstrate that for large enough values of the linear non-minimal coupling the evolution of the scalar field is reversed compared to the case of General Relativity and the field rolls `up' its potential. This leads to avoidance of the Big Crunch singularity that would occur in the context of General Relativity. In section IV we conclude and discuss extensions of the present study.

\section{Cosmic Dynamics}
The cosmological dynamics of scalar-tensor cosmological models is determined by the Lagrangian
density of the form (eg  \cite{scal-tens-quint-Capozziello:2007iu})
\be
{\cal L}=\frac{F(\phi)}{2}~R - \frac{1}{2}~g^{\mu\nu}
\partial_{\mu}\phi\,
\partial_{\nu}\phi
- V(\phi)  + {\cal L}_{\rm m}[\psi_{\rm m}; g_{\mu\nu}]\
\label{action-sctens} \ee where ${\cal L}_{\rm m}[\psi_{\rm m};
g_{\mu\nu}]$ describes matter as a
pressureless perfect fluid and we have set $8\pi G\!=\!c\!=\!1$.

The $F(\phi)$ function expresses the non-minimal coupling of the field $\phi$ with gravity and is such that when $F\!=1$ the action reduces to the usual one in General Relativity. In what follows we investigate the cosmic dynamics in the context of scalar-tensor quintessence assuming a linear potential $V(\phi)$ of the form (\ref{linpot-def}) and a linear non-minimal coupling
\be
F(\phi)=1-\lambda\phi,\label{Fphi-linanz}
\ee
where $\lambda$ is a parameter. We focus on the asymptotic future cosmological evolution of this class of models and the possible avoidance of the future singularity for some values of the parameters $s$ and $\lambda$.

Variation of the action corresponding to (\ref{action-sctens}) in the context of a flat FRW metric with pressureless matter fluid leads to the dynamical equations
\begin{gather}
3F H^2 = \rho_m +\frac{{\dot \phi}^2}{2}+V-3H{\dot F} \label{h2dyneq} \\
\ddot{\phi}+3\frac{\dot{a}}{a}\dot{\phi}-3F_\phi\left(\frac{\ddot{a}}{a}+\frac{\dot{a}^2}{a^2}\right)+V_\phi=0.\label{ddotphi-sctens}\\
-2F\left(\frac{\ddot{a}}{a}-\frac{\dot{a}^2}{a^2}\right)=\rho_m+{\dot \phi}^2+{\ddot F}-H{\dot F} \label{ddota-sctens}
\end{gather}

After rescaling with the present day Hubble parameter $H_0$ (setting $H={\bar H} H_0$, $t=\bar t/H_0$, $V={\bar V} H_0^2$ and $\rho_m={\bar \rho}_m H_0^2$ ) we obtain the density parameters for matter and dark energy from eq. (\ref{h2dyneq}) as
\ba
\Omega_m &=&\frac{{\bar \rho}_m}{3F{\bar H}^2 }\Rightarrow  \Omega_{0m}=\frac{{\bar \rho}_{0m}}{3F_0 } \label{matdenspar}\\
\Omega_\phi &=& 1-\Omega_m=\frac{1}{3F{\bar H}^2}\left(\frac{{\dot \phi}^2}{2}+{\bar V}\right) - \frac{{\dot F}}{{\bar H}F} \Rightarrow \nn \\ \Omega_{0\phi}&=&\frac{1}{3F_0} \left(\frac{{\dot \phi_0}^2}{2}+{\bar V_0}\right) - \frac{{\dot F_0}}{F_0}\label{dedenspar}
\ea
where the index $_0$ defines the values of the corresponding quantities at the present time. In what follows we use the rescaled dimensionless quantities and omit the bar in $\bar H$ and $\bar V$.

Using eqs. (\ref{ddota-sctens}) and (\ref{h2dyneq}) we obtain the dynamical equation for the scale factor as
\be
\frac{\ddot a}{a}=-\frac{\Omega_{0m}F_0}{2 a^3 F}-\frac{{\dot \phi}^2}{3F}+\frac{V}{3F}-\frac{H{\dot F}}{2F}-\frac{{\ddot F}}{2F} \label{ddota2}
\ee

\section{Numerical Solution}
We now solve the coupled system of the cosmological dynamical equations for the scalar field and for the scale factor (\ref{ddota2}) and (\ref{ddotphi-sctens}). We assume $\Omega_{0m}=0.3$ and initial conditions deep in the matter era ($t\!\ll\! t_0$) when the scalar field is assumed frozen at $\phi(t_i)\!=\!\phi_i$ (${\dot \phi}(t_i)=0$) due to cosmic friction in the context of thawing\cite{thawing-Caldwell:2005tm,thawing-lin-Scherrer:2007pu} scalar-tensor quintessence\cite{ext-quint-Nesseris:2006hp,scal-tens-cross-Perivolaropoulos:2005yv,phant-cross-Nesseris:2006er}. At that time the dynamical equation (\ref{ddota2}) reduces to
\be
\frac{\ddot a}{a}=-\frac{\Omega_{0m}F_0}{2 a^3 F_i} \label{ddotmatera}
\ee
where $F_i\equiv F(\phi_i)=1-\lambda \phi_i$. Eq. (\ref{ddotmatera}) leads to the initial conditions for the scale factor
\ba
a(t_i)&=&\left(\frac{9F_0}{4F_i}\,\Omega_{0m}\right)^{1/3}t_i^{2/3} \label{incond1}\\
{\dot a}(t_i)&=&\frac{2}{3}\left(\frac{9F_0}{4F_i}\,\Omega_{0m}\right)^{1/3}t_i^{-1/3} \label{incond2}
\ea

In order to solve the system (\ref{ddota2}) and (\ref{ddotphi-sctens}) with the above initial conditions we tune self-consistently the values of $\phi_i$ and $F_0\equiv F(\phi(t_0))=1-\lambda \phi_0$ so that the following consistency conditions are simultaneously satisfied  at the present time:
\ba
a(t_0)&=&1 \label{a01} \\
H(t_0)&=&1 \label{h01} \\
\Omega_{0\phi}&=&0.7 \label{om0fi07}\\
F(\phi(t_0))&\equiv & 1-\lambda \phi(t_0)=F_0 \label{f0f0}
\ea
In practice we define $t_0$ as the time when the scale factor is $a=1$ and then tune $\phi_i$ and $F_0$ in eq. (\ref{ddota2}) and in the initial conditions (\ref{incond1}), (\ref{incond2}) so that equations (\ref{h01}), (\ref{om0fi07}), (\ref{f0f0}) are satisfied in the numerical solution.

\begin{figure}[!t]
\centering
\includegraphics[width=\columnwidth,height=6cm]{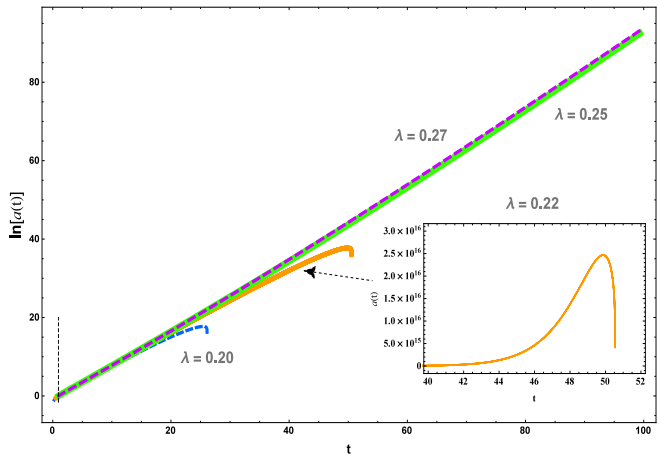}
\caption{The collapse of the scale factor for representative quintessence with linear potential ($s\!=\!1$) and various values of slopes $\lambda$. The plot is logarithmic and its linear nature indicates that for $\lambda>\lambda_{crit}\simeq 0.24$ the universe reaches a deSitter evolution. The present time corresponds to $t_0=0.96$. The magnified plot shows more clearly the evolution of the scale factor towards the Big Crunch singularity for $\lambda<\lambda_{crit}$.}\label{sc-fac}
\end{figure}

\begin{figure}[!t]
\centering
\includegraphics[width=\columnwidth,height=6cm]{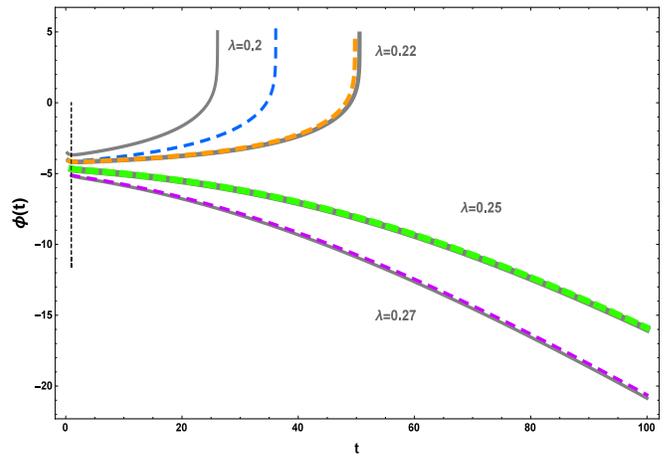}
\caption{The evolution of the scalar field for $s=1$ and values of $\lambda$ above and below the critical values $\lambda_{crit}(s=1)=0.24\pm 0.01$. For $\lambda<\lambda_{crit}$ the field rolls towards larger values down its potential $V=-s\phi$ (upper curves). Thus the potential becomes negative with strongly attractive gravity leading to a Big Crunch singularity. For stronger non-minimal coupling $\lambda>\lambda_{crit}$ (lower curves) the field evolves towards negative values (up its potential). The positive potential energy of the field eventually dominates and its repulsive gravity leads to eternal expansion avoiding the Big Crunch singularity. The solid lines correspond to initial conditions ${\dot \phi}_i=0$ while the dashed lines correspond to ${\dot \phi}_i=15$.}
\label{sc-field}
\end{figure}

In Fig. \ref{sc-fac} we show the evolution of the scale factor for various values of the parameter $\lambda$ and $s=1$. For small values of $\lambda$ the evolution of the scale factor is similar to the one anticipated in the minimally coupled case \cite{lin-neg-Perivolaropoulos:2004yr}. Initially the universe expands with a late time acceleration but soon after the field potential develops to negative values, the scalar field gravitational interaction becomes strongly attractive and the scale factor collapses to a singularity (doomsday).  However, for values of $\lambda$ larger than a critical value $\lambda_{crit}$, the nonminimal coupling becomes important and the dynamics of the scalar field change at late times due to the term proportional to $\lambda$ in eq.  (\ref{ddotphi-sctens}). Instead of rolling down the potential towards larger field values, it starts rolling up its potential towards smaller (negative) field values as dictated by its non-minimal coupling to the metric (lower curves in Fig. \ref{sc-field}).
This is shown in Fig \ref{sc-field}  where we present the time evolution  of the scalar field for values of $\lambda$ below and above the critical value which for $s=1$ is approximately $\lambda_{crit} \simeq 0.24$. The evolution of the scalar field potential is trivially obtained from Fig. \ref{sc-field} using eq. (\ref{linpot-def}) ($V(\phi)=-\phi$). A result of this reversal of the potential energy evolution of the scalar field towards positive values is the continuation of the accelerating expansion and the avoidance of the collapse towards a singularity for $\lambda > \lambda_{crit}$ as shown in Fig. \ref{sc-field}. We have verified this result by extending the numerical evolution of the scale factor to significantly larger times than shown and also using different initial conditions for the initial field time derivative (see Fig. \ref{sc-field}).

We have focused on the class of initial conditions that can reproduce approximately our universe (where the relative dark energy density $\Omega_{0\phi}$ is about 0.7 (eq. (\ref{om0fi07})) and its future evolution. This constraint fixed the initial value of the scalar field in accordance with eqs (\ref{a01})-(\ref{f0f0}) but leaves freedom in selecting its initial time derivative.  We have explored a range of the initial derivative $\dot \phi_i$ and found that even though it can mildly affect the details of the future evolution of the scalar field it does not affect  the critical values of $\lambda$. This is also demonstrated in  Fig. \ref{sc-field} where we show the field evolution for $\dot \phi_i=0$ and $\dot \phi_i=15$.

Thus, the existence of a critical value of $\lambda$ is demonstrated in Fig. \ref{sc-field} where the evolution of the scalar field is shown to change in a dramatic manner when $\lambda$ crosses the critical value.

In view of the fact that the doomsday singularity is avoided for $\lambda > \lambda_{crit}$ we wish to address the following questions:
\begin{itemize}
\item
What is the actual future cosmological evolution for $\lambda > \lambda_{crit}$?
\item
What is the dependence of $\lambda_{crit}$ on the slope $s$ of the potential?
\end{itemize}

In the context of addressing the first question we derive the evolution of the scalar field dark energy (DE) equation of state parameter from the generalized field equations (\ref{h2dyneq}) and (\ref{ddota-sctens}). These can be rewritten as \cite{scal-tens-quint-Capozziello:2007iu}
\begin{gather}
3F_0H^2=\rho_{DE}+\rho_m \label{h2-sctens-f0}\\
-2F_0\dot{H}=\rho_{DE}+p_{DE}+\rho_m \label{doth-sctens-f0}
\end{gather}
where we have set
\begin{gather}
\rho_{DE}=\frac{1}{2}\dot{\phi}^2+V-3H^2\left(F-F_0\right)-3H\dot{F} \label{rhode-sctens}\\
p_{DE}=\frac{1}{2}\dot{\phi}^2-V+\ddot{F}+2H\dot{F}+\left(2\dot{H}+3H^2\right)\left(F-F_0\right)\label{pde-sctens}
\end{gather}

The function $\rho_{DE}$ defined in this way can be shown to satisfy the usual energy conservation equation:
\begin{equation}
\dot{\rho}_{DE}+3H\left(\rho_{DE}+p_{DE}\right)=0,\label{encons-sctens}
\end{equation}
Using  eqs (\ref{rhode-sctens}) and (\ref{pde-sctens}) it is straightforward to show that the scalar field dark energy (DE) equation of state parameter $w_{DE}=\frac{p_{DE}}{\rho_{DE}}$ may be written as
\begin{equation}
w_{DE}=-1+\frac{\dot{\phi}^2+\ddot{F}-H\dot{F}-2\dot{H}\left(F_0-F\right)}{\frac{1}{2}\dot{\phi}^2+V-3H^2\left(F-F_0\right)-3H\dot{F}} \label{wde-sctens}
\end{equation}
Using the dynamical equations (\ref{ddota-sctens}), (\ref{h2dyneq}), $w_{DE}$ may be connected with the observable Hubble parameter as
\begin{equation}
w_{DE}=-\frac{3H^2(z)-(1+z)\left(\mathrm{d}H^2(z)/\mathrm{dz}\right)}{3H^2(z)-3\,\Omega_{0,m}\left(1+z\right)^3}
\label{wde-sctens2}
\end{equation}

\begin{figure}[!t]
\centering
\includegraphics[width=\columnwidth,height=6cm]{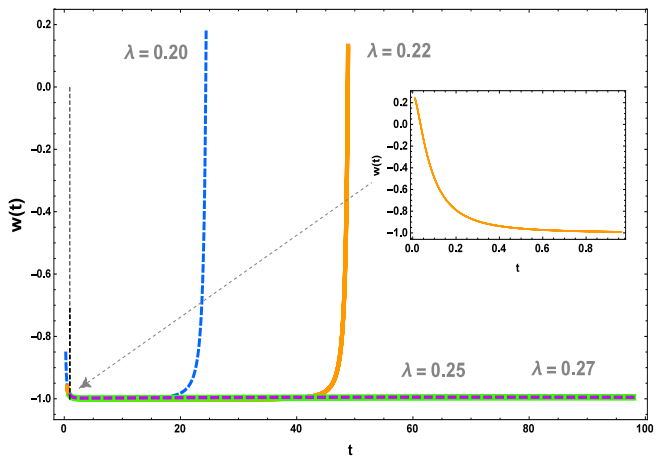}
\caption{The evolution of the equation of state parameter for $s=1$ and values of $\lambda$ above and below the critical values $\lambda_{crit}(s=1)=0.24\pm 0.01$. The present time $t_0$ (obtained by demanding $\Omega_m(t_0)=0.3$) corresponds to $t=t_0=0.96$. For $\lambda >\lambda_{crit}\simeq 0.24$ the universe continues its expansion reaching a deSitter phase since $w_{DE}=-1$. For $\lambda <\lambda_{crit}$ the dark energy equation of state $w_{DE}$ becomes positive and diverges leading to infinite attractive gravity and a Big Crunch singularity at a finite future time which depends on the values of the parameters $\lambda$ and $s$.}\label{eqstat}
\end{figure}

The equivance of eqs (\ref{wde-sctens}) and (\ref{wde-sctens2}) was also verified numerically as a test of the validity of our analysis.
We obtain the evolution of the equation of state parameter (\ref{wde-sctens}) by solving the system of dynamical equations  (\ref{ddota2}) and (\ref{ddotphi-sctens}) as described above. This evolution is shown in Fig. \ref{eqstat} for $s=1$ and four values of $\lambda$ chosen below and above the critical value which for $s=1$ is $\lambda_{crit}\simeq 0.24$. For $\lambda<\lambda_{crit}$ $w_{DE}$ reaches the value $-1$ corresponding to domination of the potential energy or an effective  cosmological constant but after the potential and the effective cosmological constant becomes negative, $w_{DE}$ departs from the value $-1$ and diverges towards positive values. In contrast for $\lambda>\lambda_{crit}$, $w_{DE}$ remains at the value $-1$ and the universe reaches a deSitter phase.

\begin{figure}[!t]
\centering
\includegraphics[width=\columnwidth,height=6cm]{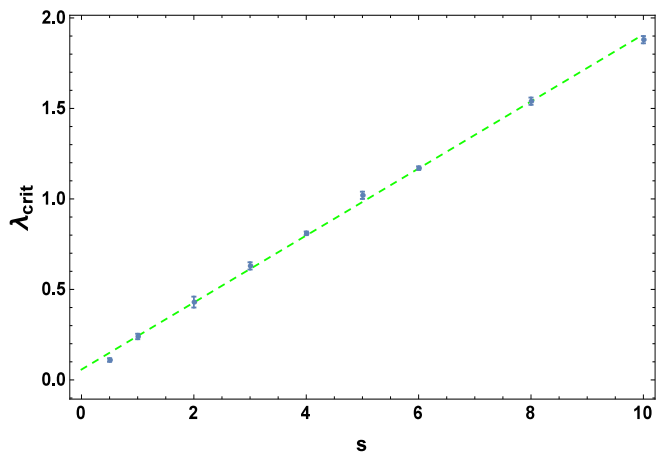}
\caption{The critical value of $\lambda$ for various values of the slope $s$ of the linear potential. The dependence of $\lambda_{crit}$ on $s$ is approximately linear. For $\lambda > \lambda_{crit}(s)$ the Big Crunch doomsday singularity is avoided.}\label{lcrit}
\end{figure}

We have repeated the above analysis for various values of the slope $s$ of the potential in order to obtain the $\lambda_{crit}(s)$ (Fig. \ref{lcrit}). For small values of the slope $s$, the dynamics leading to a singularity (doomsday) as determined by the potential $V(\phi)$ can be reversed by a small value of the non-minimal coupling parameter $\lambda$. The required value of $\lambda$ for reversal of the doomsday dynamics increases almost linearly with $s$ as shown in Fig. \ref{lcrit}.

\section{Discusssion-Conclusion}
In conclusion, we have shown that the scale factor Big Crunch singularity that is generically present in quintessence with linear potentials  is avoided in scalar-tensor quintessence for a range of values of the non-minimal coupling. This is due to the modified dark energy dynamics induced by the non-minimal coupling present in scalar-tensor quintessence.

Interesting extensions of the present study include the following:
\begin{itemize}
\item
Observational constraints in the parameter space $(\Omega_{0m},s,\lambda)$ to be imposed using recent cosmological data \cite{data-planck-Ade:2015xua,data-snia-comb-Betoule:2014frx,data-sdss-Dawson:2015wdb,data-bao-Aubourg:2014yra} and compared with the parameter range corresponding to the avoidance of the doomsday derived in the present study and the consistency with cosmological and solar system constraints\cite{cosmo-constr-G-Umezu:2005ee,constr-sc-tens-Will:2005va,cosmo-bounds-G-Gaztanaga:2001fh,cosmo-constr-sctens-Avilez:2013dxa}. A chameleon mechanism could significantly relax such constraints\cite{chameleon-Khoury:2003rn}.
\item
A possible analytical derivation \cite{anal-quint-Harko:2013gha} of the numerically derived values of $\lambda_{crit}(s)$ which appears to be an approximately linear function.
\item
Investigation of alternative forms of quintessence potentials that lead to collapsing singularities and the possible avoidance of these singularities in the context of scalar-tensor quintessence.
\end{itemize}

In order to obtain a qualitative estimate of the range of parameter values that are consistent with cosmological observations we plot the scalar field equation parameter obtained from eq. (\ref{wde-sctens}) as a function of redshift in the range $z\in [0,2]$. We consider several values of the parameter pair $(s,\lambda)$ and fit the numerically obtained $w(z)$ with the Chevallier-Polarski-Linder (CPL) parametrization \cite{Chevallier:2000qy,Linder:2002et}
\be
w(z)=w_0 +w_1 \frac{z}{1+z}
\label{cplanz}
\ee
thus obtaining a best (least squares) fit parameter pair $(w_0,w_1)$ for each pair $(s,\lambda)$.
Comparing with the current cosmological constraints  \cite{Hazra:2013dsx}
\be
w_0=-1.005^{+0.15}_{-0.17}, w_1=-0.48^{+0.77}_{-0.54}
\label{wconstr}
\ee
we may obtain an estimate for the range of $(s,\lambda)$ that is consistent with observations. A parameter estimation using direct fit of the numerically obtained Hubble parameter to Type Ia supernova (SnIa), Baryon Acoustic Oscillations (BAO) and Cosmic Microwave Background (CMB)data would be more quantitative and accurate but is beyond the goals of the present study.

\begin{figure*}[ht]
\centering
\begin{center}
$\begin{array}{@{\hspace{-0.10in}}c@{\hspace{0.0in}}c}
\multicolumn{1}{l}{\mbox{}} &
\multicolumn{1}{l}{\mbox{}} \\ [-0.2in]
\epsfxsize=3.3in
\epsffile{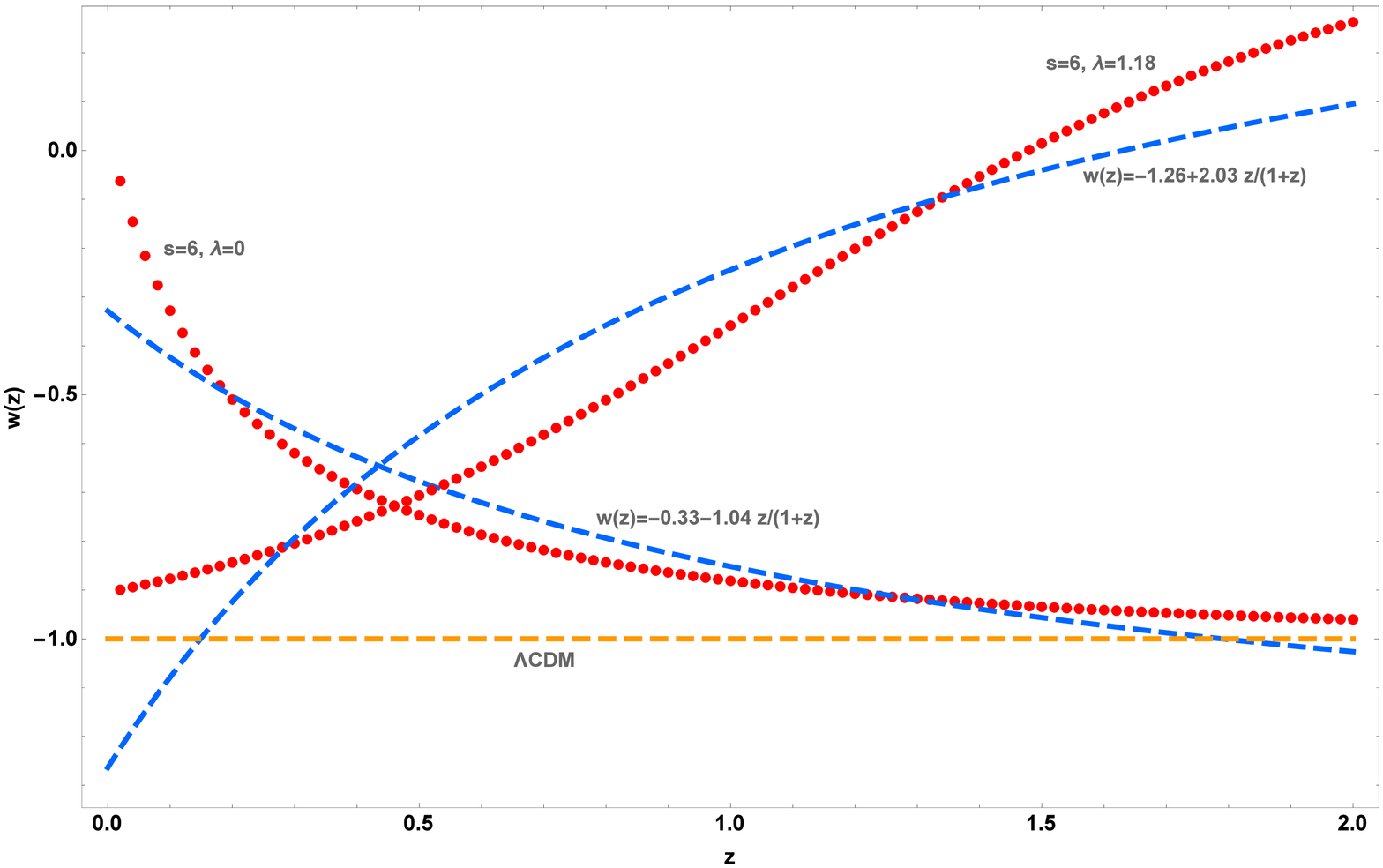} &
\epsfxsize=3.3in
\epsffile{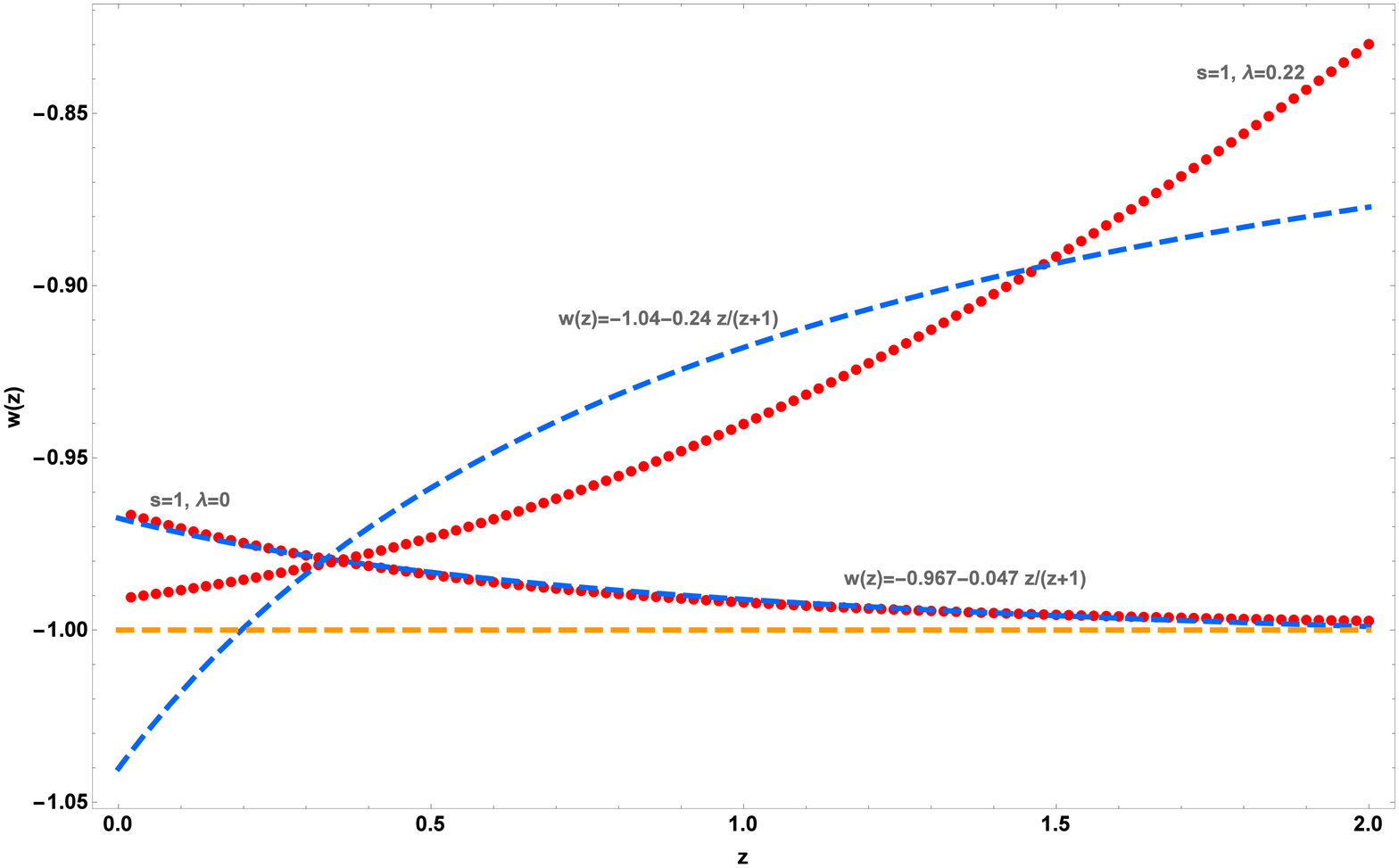} \\
\end{array}$
\end{center}
\vspace{0.0cm}
\caption{\small a: The numerical evolution of the scalar field equation of state parameter $w(z)$ from $z=2$ up to the present time ($z=0$) (red dotted lines) superposed with their corresponding best fits corresponding to the CPL parametrization $w(z)=w_0 +w_1 \frac{z}{1+z}$ \cite{Chevallier:2000qy,Linder:2002et} (blue dashed lines). This plot corresponds to parameter values $s=6, \lambda=0$ (best fitted by $w_0=-0.33, w_1=-1.04$) and $s=6, \lambda=1.18$ (best fitted by $w_0=-1.26, w_1=2.03$). These best fit CPL parameter values are not consistent with current cosmological constraints $w_0=-1.005^{+0.15}_{-0.17}, w_1=-0.48^{+0.77}_{-0.54}$ \cite{Hazra:2013dsx}. b: Similar to (a) for parameter values $s=1, \lambda=0$ (best fitted by $w_0=-0.967, w_1=-0.047)$ and $s=1, \lambda=0.22$ (best fitted by $w_0=-1.04, w_1=-0.24$). These parameter values are consistent with current cosmological constraints at $1\sigma$ level.}
\label{wfit}
\end{figure*}
In Fig. \ref{wfit} we show the numerically obtained form of $w(z)$ for four parameter pairs $(s,\lambda)$. The corresponding CPL parametrization best fits are also shown. On the left we show parameter pairs whose best fit CPL parameters are well outside the observationally obtained $1\sigma$ range while the right panel of Fig. \ref{wfit} shows examples of $w(z)$ for parameters that are within the allowed $1\sigma$ range. Notice that for $\lambda \neq 0$ the scalar-tensor quintessence model behaves like freezing quintessence (approaches $w=-1$ at $z=0$) while for $\lambda = 0$ the behavior is that of thawing quintessence. It is also clear that the CPL parametrization is much more efficient in fitting a thawing quintessence behavior than fitting the freezing quintessence type of evolution. This observation implies that a different parametrization would be more efficient in representing dark energy models whose equation of state parameter approaches smoothly the value $-1$ at recent times (like freezing quintessence).

In Table I we show the best fit CPL parameter values for a set of parameters $(s,\lambda)$. Comparison with the constrain equation (\ref{wconstr}) leads to a qualitative estimate of constraints on the parameters $(s,\lambda)$. Based on Table I, a qualitative rough estimate of the parameter constraints at $1\sigma$ is $s\lsim 2$ $\lambda \lsim 0.5$

\vspace{0pt}
\begin{widetext}
\begin{table}[b!]
\begin{center}
\begin{minipage}{0.88\textwidth}
\caption{The best fit $(w_0,w_1)$ pairs corresponding to $(s,\lambda)$ pairs. The $\lambda_{crit}$ for each $s$ may seen in Fig. \ref{lcrit}. \label{table1}}
\end{minipage}\\
\begin{tabular}{c|c|c|c|c|c|c}
\hline
\hline\\
\bf{$\lambda_{crit}$ , $s$}&$s=1$ & $s=2$ & $s=4$ &$s=6$ &$s=8$ &$s=10$ \\
\\\hline \vspace{0pt}\\
\bf{$\lambda=0$} & (-0.96,-0.05)  \hspace{7pt}& (-0.88,-0.18)  \hspace{7pt}& (-0.61,-0.59)   \hspace{7pt}& (-0.32,-1.05)   \hspace{7pt}&(-0.09,-1.45)   \hspace{7pt}&(0.13,-1.82)   \hspace{7pt} \\
\bf{$\lambda=\lambda_{crit}(s)$ } & (-1.04,0.24)   \hspace{7pt}& (-1.13,0.81)   \hspace{7pt}&  (-1.26,1.8)   \hspace{7pt}& (-1.26,2.05)   \hspace{7pt}&(-1.23,2.04)   \hspace{7pt}&(-1.21,-1.97)   \hspace{7pt} \\
 \hline \hline
\end{tabular}
\end{center}
\end{table}
\end{widetext}

A possible approach for demonstrating analytically the existence of critical values for the parameter $\lambda$ could be obtained by deriving an effective evolution equation for the scalar field $\phi$. As a first step in that direction equations (\ref{ddota-sctens}) and (\ref{h2dyneq}) may be used in eq. (\ref{ddotphi-sctens}) to obtain the effective evolution equation for $\phi$ of the form

\ba
&&{\ddot \phi}\left(1+\frac{3\lambda^2}{2F}\right)+3 H \left(1+\frac{3 \lambda^2}{2F}\right){\dot \phi}-\frac{\lambda}{2F} {\dot \phi}^2= \nn \\ && =s+\frac{2 \lambda s \phi}{F} - \lambda \frac{\Omega_{0m}}{2 a^3}\frac{F_0}{F}
\label{ddotphi-dec}
\ea

This is a Rayleigh equation which has some similarities with the standard forced-damped oscillator.  However, there are important differences like the presence of the non-linear term of the field derivative which complicate the analysis and do not allow the use of the sign of the `force' term as a qualitative simple indicator of the dynamics. Nevertheless, equation (\ref{ddotphi-dec}) could be a useful starting point for the analytical understanding of the existence of critical points $\lambda_{crit}$ since it demonstrates that for negative initial values of the scalar field (positive potential energy) and large enough values of $\lambda$ the sign of the effective force changes and drives the scalar field towards negative values (up its potential). A detailed analysis along these lines is outside the scope of the present study and is postponed for a future analysis.


The Mathematica files that were used for the numerical analysis and the production of the figures may be downloaded from \href{https://drive.google.com/file/d/0B7rg6X3QljQXRzVDTXNSRTA3X2s/view}{this url}.

\raggedleft
\bibliography{biblio}



\end{document}